%% file: refgen.tex
\documentclass[journal]{IEEEtran}
\usepackage[utf8]{inputenc}
\usepackage{graphicx}
\usepackage{cite}
\usepackage{amsmath,amssymb,amsfonts}
\usepackage{textcomp}
\usepackage{subfigure}
\usepackage{newtxtext} 
\usepackage{newtxmath}
\usepackage{bm}
\usepackage{booktabs}
\usepackage{flushend}
\usepackage{url}
\usepackage{hyperref}

\input{vardefs}

\graphicspath{{figs/}} 

\begin{document}

\title{Decoupled Online Feedforward Generation\\ of Optimal References for Saturated\\ Synchronous Machine Drives}

\author{
	\vskip 1em
	Mikko~Sarén, Hannu~Hartikainen, Antti~Piippo, and~Marko~Hinkkanen, \emph{Fellow, IEEE}
	\thanks{
	This work has been submitted to the IEEE for possible publication. This arXiv version includes minor bug fixes and language improvements compared with the version submitted to the journal. Copyright may be transferred without notice, after which this version may no longer be accessible.
    
	This work was supported by ABB Oy as well as by the Research Council of Finland through the \textit{Center of Excellence in High-Speed Electromechanical Energy Conversion Systems}. The authors acknowledge the use of EPE infrastructure of Aalto School of Electrical Engineering.

	Mikko Sarén, Hannu Hartikainen, and Marko Hinkkanen are with Aalto University, Espoo, Finland (e-mail: mikko.saren@aalto.fi, hannu.1. hartikainen@aalto.fi, marko.hinkkanen@aalto.fi).

	Antti Piippo is with ABB Oy, Drives, 00380 Helsinki, Finland (e-mail: antti.piippo@fi.abb.com).}
}
\maketitle	

\begin{abstract}
This paper presents a modular method for generating reference signals online for saturable synchronous machine drives. The method dynamically generates optimal references without precomputed lookup tables, following the maximum-torque-per-ampere (MTPA) trajectory while respecting maximum-torque-per-volt (MTPV), current, and voltage limits. The proposed tracking laws are formulated to yield exact, decoupled first-order error dynamics, ensuring predictable tracking responses and simplifying system tuning. The algorithm requires only the forward flux map, thereby eliminating the need for current-map inversion. By operating in a feedforward manner, the method ensures noise-free reference signals and structural separation from the feedback control. Both simulation and experimental results are presented, demonstrating that the proposed method achieves dynamic and steady-state performance on par with conventional lookup-table-based approaches, while avoiding the need for precomputed reference tables.
\end{abstract}


\section{Introduction}
\IEEEPARstart{P}{ermanent-magnet} (PM) and reluctance synchronous machines are utilized in demanding applications ranging from electric vehicles to industrial servo drives. Achieving maximum efficiency and full utilization of inverter capacity in these drives requires precise reference generation strategies that minimize resistive losses while rigorously respecting system constraints. In particular, reference signals must satisfy the maximum-torque-per-ampere (MTPA) condition and, during high-speed operation, comply with maximum-torque-per-volt (MTPV), current, and voltage limits. However, magnetic saturation introduces significant nonlinearities in the flux-current relationship, complicating accurate real-time tracking of these optimal conditions. 

Traditionally, optimal reference signals are generated using precomputed lookup tables~\cite{Mey2006, Che2010, Awa2018,Dia2022}. While straightforward to execute, creating and implementing these lookup tables is cumbersome and potentially error-prone. Furthermore, static lookup tables cannot seamlessly accommodate dynamic changes in the machine. If the magnetic model or PM-flux linkage is updated online---for example, due to thermal variations---the precomputed reference tables become immediately inconsistent with the active control algorithm, leading to sub-optimal performance.

To avoid the rigid nature of lookup tables, explicit analytical solutions have been developed to calculate optimal current references in closed form. For example, by assuming a linear magnetic model, the MTPA problem can be formulated as a fourth-order polynomial and solved in real time using algebraic root-finding algorithms, such as Ferrari's method~\cite{Jun2013}. Other mathematical formulations model the operating constraints as quadric surfaces to derive instantaneous analytical references~\cite{Eld2017}. While these algebraic approaches are computationally deterministic and bypass the need for iterative solvers, their accuracy deteriorates significantly under heavy magnetic saturation. Deriving exact closed-form solutions that incorporate highly nonlinear, cross-saturated arbitrary flux maps remains mathematically prohibitive.

To avoid nonlinear magnetic modeling entirely, parameter-independent strategies such as extremum seeking control and high-frequency signal injection have been widely investigated~\cite{Bol2011, Li2019, Li2024}. By injecting an active probing signal into the stator current or voltage and evaluating the resulting power or torque ripple, these methods search for the MTPA operating point under saturation and thermal variation. However, physical signal injection introduces torque ripple, acoustic noise, and additional losses. While virtual signal injection techniques~\cite{Sun2015, Sun2017, Wan2018} reduce these physical effects, all search-based methods inherently suffer from slow transient responses. 

To account for magnetic nonlinearity without relying on search-based algorithms, online feedforward reference generation methods based on numerical optimization have been proposed~\cite{Jeo2006, Bon2018, Kim2019}, and subsequently extended to include MTPV limits~\cite{Xia2020}. However, these approaches typically rely on iterative solvers executing at every sampling instant. The heavy computational burden of solving nonlinear optimization problems in real time limits their practical deployment. 

As a computationally lighter alternative, online feedback-based reference-generation methods have been proposed~\cite{Var2021b, Var2021, Var2022}. Unlike purely feedforward methods, these approaches compute part of the reference signals from measured or estimated drive variables. This reduces the need for online optimization, but it also creates a feedback path from the closed-loop system to the reference generator. As a result, measurement noise and estimation errors can propagate directly to the generated references. In addition, the dynamics of the reference generator may interact with the main feedback controller, which can complicate stability analysis and tuning.

A recent feedforward approach avoids iterative optimization and feedback noise~\cite{Har2025}. However, it requires implementations of both forward flux and inverse current maps. While references are tracked online, the underlying laws yield coupled error dynamics without an analytical solution for the gains. This severely complicates tuning.

To overcome these limitations, this paper proposes an online feedforward method for generating reference signals for saturable synchronous machine drives. The method generates the references dynamically, without precomputed lookup tables. It follows the MTPA trajectory and seamlessly respects the MTPV, current, and voltage limits over the operating range. The key contributions of this paper are:
\begin{itemize}
    \item Decoupled error dynamics: Unlike earlier tracking algorithms, the proposed tracking laws are formulated to yield exact, decoupled first-order error dynamics. As a result, the transient response of each tracking subsystem is precisely known and can be tuned directly using its desired bandwidth.
    \item Direct flux-map formulation: The proposed formulation requires only the forward flux map, entirely bypassing the need for inverse current maps. This simplifies the real-time implementation and reduces computational overhead.
    \item Experimental validation: The proposed dynamic reference generator is validated experimentally on a 5.6-kW PM synchronous reluctance machine drive, demonstrating its practical viability and high-performance transient capability.
\end{itemize}
The proposed reference generator operates in a feedforward manner and is structurally separated from the main feedback controller. Therefore, measurement noise from the feedback signals is not directly fed into the generated references. This separation also prevents the reference-generator dynamics from forming an additional feedback path through the drive control system, which simplifies tuning and stability analysis.

The framework is modular and can be used with different control strategies. In flux-vector control and volts-per-hertz (V/Hz) control, the generated flux and torque references can be used directly. In current-vector control, an additional dynamic subsystem converts these scalar references into current references. Thus, only minor modifications are needed when the method is applied with different control structures.

This article is organized as follows. Section~\ref{sec:model} reviews the fundamental machine model and the mathematical formulations of the MTPA and MTPV conditions. Section~\ref{sec:ref_gen} details the proposed optimal reference generation framework and tracking laws. Section~\ref{sec:results} presents the simulation and experimental results, followed by concluding remarks in Section~\ref{sec:conclusion}.

\section{Machine Model} \label{sec:model} 

\subsection{Fundamentals}
A saturable synchronous machine is modeled in rotor coordinates using per-unit quantities. Real space vectors are employed, e.g., the stator current is $\is = [\isd, \isq]^\T$ and the stator flux linkage is $\psis = [\psisd, \psisq]^\T$, with similar notation for other variables. The corresponding state equations are
\begin{align} \label{eq:dpsis}
    \frac{\D \psis}{\D t} &= \us - \Rs \is - \omegam \J \psis \\ \label{eq:dthetam}
    \frac{\D \thetam}{\D t} &= \omegam 
\end{align}
where $\us$ is the stator voltage, $\Rs$ is the stator resistance, $\omegam$ is the rotor angular speed, $\thetam$ is the rotor angle, and $\J = [\begin{smallmatrix} 0 & -1 \\ 1 & 0 \end{smallmatrix}]$ is the orthogonal rotation matrix. 

The magnetic behavior is described by a nonlinear conservative magnetic model. Expressed in terms of the magnetic energy state function $W(\psis)$, the stator current is
\begin{align} \label{eq:magnetic_model}
    \is = \left[\frac{\partial W{(\psis)}}{\partial \psis}\right]^\T
\end{align}
and the electromagnetic torque is given by
\begin{align} \label{eq:tauM}
    \TM = \is^\T\J\psis 
\end{align}
Therefore, in this energy-based formulation, both the stator current $\is(\psis)$ and the torque $\TM(\psis)$ are functions of the stator flux linkage. 

Alternatively, the dual magnetic model can be expressed in terms of the co-energy state function $W'(\is)$ as
\begin{align} \label{eq:is_co}
    \psis = \left[\frac{\partial W'(\is)}{\partial \is}\right]^\T 
\end{align}
Hence, in this formulation, the stator flux linkage $\psis(\is)$ is a function of the stator current. Consequently, substituting~\eqref{eq:is_co} into~\eqref{eq:tauM} shows that the electromagnetic torque also becomes a function of the stator current, $\TM(\is)$.

The standard linear magnetics model is obtained as a special case with $W'(\is) = \frac{1}{2} \Ld \isd^2 + \frac{1}{2} \Lq \isq^2 + \abspsif \isd$, where $\Ld$, $\Lq$, and $\abspsif$ are constant parameters. This choice yields the linear relationship $\psis(\is) = [\begin{smallmatrix} \Ld & 0 \\ 0 & \Lq \end{smallmatrix}]\is + [\begin{smallmatrix} \abspsif \\ 0 \end{smallmatrix}]$. In this paper, this constant-parameter relation is replaced by the general nonlinear flux map $\psis(\is)$. 

\begin{figure}[t] \centering
    \subfigure[]{\includegraphics[scale=.85]{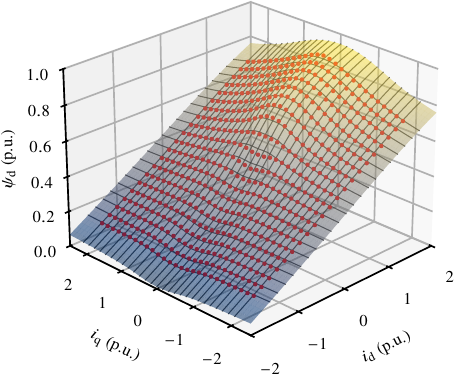}}
    \subfigure[]{\includegraphics[scale=.85]{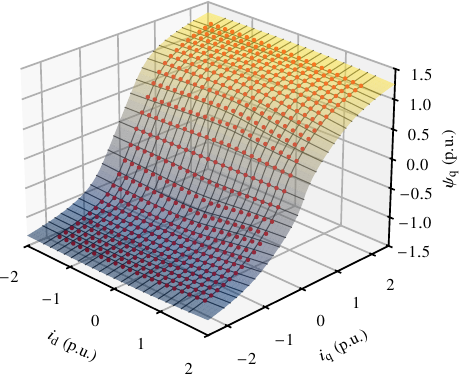}}
    \caption{Flux linkage maps for a 5.6-kW PM synchronous reluctance machine: (a) $\psisd$; (b) $\psisq$. The red points show the measured data from the constant-speed test~\cite{Arm2013}. The surfaces illustrate the analytical model that has been fitted to the measured data~\cite{Lel2024}.} \label{fig:sat1}
\end{figure}

A 5.6-kW four-pole PM synchronous reluctance machine is used as an example to visualize the nonlinear magnetic characteristics and the optimal operating loci. Fig.~\ref{fig:sat1} shows the measured flux-linkage data and the analytical saturation model fitted to these data. The rated values, parameters, and experimental setup are described in Section~\ref{sec:results}.

\begin{figure}[t] \centering
    \subfigure[]{\includegraphics[scale=.9,trim=0cm .3cm 0cm .1cm]{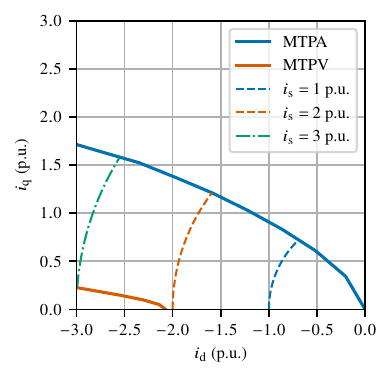}}
	\subfigure[]{\includegraphics[scale=.9,trim=0cm .3cm 0cm .1cm]{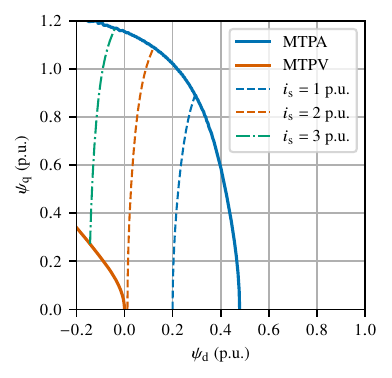}}
    \caption{MTPA, MTPV, and constant current loci presented in: (a) current plane; (b) flux linkage plane.} \label{fig:mtpa_mtpv}
\end{figure} 

\begin{figure}[t] \centering 
	\includegraphics[scale=.9,trim=0cm .5cm 0cm .1cm]{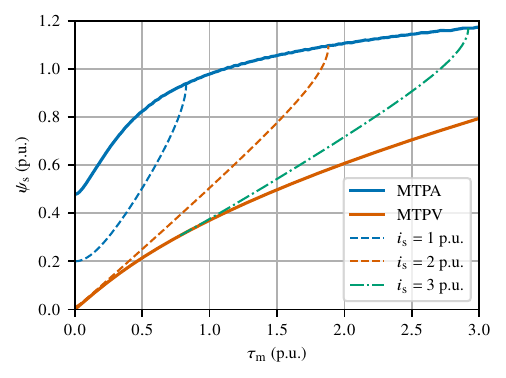}
    \caption{Flux linkage magnitude versus torque showing the MTPA, MTPV, and constant current loci, corresponding to Fig.~\ref{fig:mtpa_mtpv}.} \label{fig:mtpa_mtpv_for_ctrl}
\end{figure} 

\begin{figure*}[t] \centering 
    \includegraphics{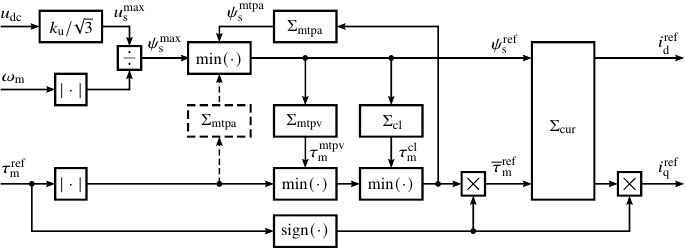}
    \caption{Block diagram of the feedforward reference generation architectures. The conventional lookup-table-based method utilizes static 1D lookup tables ($\Gmtpa$, $\Gmtpv$, and $\Glim$) alongside one or two 2D lookup tables ($\Gcur$). The lookup-table route for MTPA is marked by a dashed line. The proposed method replaces these with dynamically tracked subsystems: $\Gmtpa$, $\Gmtpv$, $\Glim$, and $\Gcur$ are employed to dynamically track the MTPA condition, the MTPV limit, the current limit, and the current reference, respectively. Note that the proposed method uses $|\TMreflim|$ instead of $|\TMref|$. The tracking system $\Gcur$ is needed only in current-vector control, not in flux-vector control.} \label{fig:ref_gen}
\end{figure*}
 
\subsection{MTPA and MTPV Conditions}
This subsection formulates the optimality conditions used by the proposed reference generator. The aim is to express the MTPA and MTPV conditions as scalar equations. These scalar equations are subsequently used as error signals in the dynamic tracking laws.

For a given torque, the MTPA condition corresponds to the operating point with the minimum stator current magnitude. Using the co-energy model in~\eqref{eq:is_co}, the gradient of the electromagnetic torque with respect to the stator current is~\cite{Var2022}
\begin{align} 
    \psia = \left(\frac{\partial \TM}{\partial \is}\right)^\T = \J\psis - \Ls \J \is
\end{align}
where the incremental inductance matrix is
\begin{align} \label{eq:Ls}
    \Ls = \frac{\partial \psis}{\partial \is}
\end{align}
Since the magnetic model is conservative, the inductance matrix is symmetric, $\Ls = \Ls^\T$. Its elements can be computed from a differentiable flux map.

At the MTPA point, the constant-torque curve is tangent to the constant-current-magnitude circle. This condition can be written compactly as
\begin{align} \label{eq:cgamma}
    \cgamma = \psia^\T\J\is = 0
\end{align}
Consequently, $\cgamma$ can be used as the scalar error signal for the MTPA condition. Its gradient with respect to the stator current is
\begin{align} \nonumber
    \phia &= \left(\frac{\partial \cgamma}{\partial \is}\right)^\T =
    -\J\psia + \frac{\partial \psia}{\partial \is}\J \is \\ \label{eq:phia}
    &= -\J\psia + (\J\Ls - \Ls\J)\J\is - \left(\frac{\partial \Ls}{\partial \is}\J\is\right)\J\is
\end{align}
The last term appears because, in a saturated machine, the incremental inductance matrix depends on the current. This term is computationally demanding. As shown in the Appendix, it can be omitted in practical implementations without changing the exact steady-state condition.

Applying the condition~\eqref{eq:cgamma}, the optimal MTPA locus of the example machine is computed and shown in Fig.~\ref{fig:mtpa_mtpv} in both the current and flux-linkage planes. Fig.~\ref{fig:mtpa_mtpv_for_ctrl} shows the corresponding relation between flux-linkage magnitude and electromagnetic torque.

The MTPV condition is obtained in a dual way. It can be interpreted as the point where the constant-torque curve is tangent to a constant-flux-magnitude contour. The gradient of the electromagnetic torque with respect to the stator flux linkage is~\cite{Var2022}
\begin{align} 
    \ia = \left(\frac{\partial \TM}{\partial \psis}\right)^\T = -\J\is + \Gs \J\psis 
\end{align}
where the incremental inverse inductance matrix is
\begin{align} \label{eq:Gs}
    \Gs = \Ls^{-1} = \left(\frac{\partial \psis}{\partial \is} \right)^{-1}
\end{align}
The MTPV condition can be expressed as
\begin{align} \label{eq:cdelta}
    \cdelta = \ia^\T\J\psis = 0
\end{align}
Its gradient with respect to the flux linkage is 
\begin{align} \nonumber
    \ja &= \left(\frac{\partial \cdelta}{\partial \psis}\right)^\T =
    -\J\ia + \frac{\partial \ia}{\partial \psis}\J \psis \\ \label{eq:ja}
    &= -\J\ia - (\J\Gs - \Gs\J)\J\psis + \left(\frac{\partial \Gs}{\partial \psis}\J\psis\right)\J\psis
\end{align}
Again, the last term is caused by magnetic saturation. As shown in the Appendix, this term can also be omitted in practical implementations without affecting the exact steady-state condition.

The MTPA and MTPV conditions have the same structure: the first is naturally written in the current plane, whereas the second is naturally written in the flux-linkage plane. However, the proposed reference generator is formulated using only the forward flux map $\psis(\is)$. Therefore, the MTPV quantities are also evaluated as functions of the stator current. For this purpose, the gradient of the flux magnitude with respect to the current is 
\begin{align}
    \lvect = \left(\frac{\partial \abspsis}{\partial \is}\right)^\T = \Ls \frac{\psis}{\abspsis} 
\end{align}
and the gradient of $\cdelta$ with respect to the current is obtained from the chain rule as
\begin{align}
    \phie = \left(\frac{\partial \cdelta}{\partial \is}\right)^\T = \Ls \ja 
\end{align}
Using the condition~\eqref{eq:cdelta}, the optimal MTPV locus of the example machine is shown in Figs.~\ref{fig:mtpa_mtpv} and~\ref{fig:mtpa_mtpv_for_ctrl}.

\section{Optimal Reference Generation} \label{sec:ref_gen}

\subsection{Conventional Lookup-Table-Based Method} 
This subsection reviews the conventional lookup-table-based implementation, which is used as the baseline for the proposed online method. Fig.~\ref{fig:ref_gen} shows a modular feedforward reference-generation architecture~\cite{Mey2006}. In the conventional implementation, the blocks $\Gmtpa$, $\Gmtpv$, and $\Glim$ are static one-dimensional (1D) lookup tables, while $\Gcur$ is implemented using one or two two-dimensional (2D) lookup tables \cite{Awa2018}.

The MTPA flux reference is obtained from  
\begin{align}
	\abspsis^\mtpa = \Gmtpa(|\TMref|) 
\end{align}
where $\TMref$ is the torque reference. Fig. \ref{fig:mtpa_mtpv_for_ctrl} shows the corresponding precomputed MTPA lookup table for the example machine. The active stator flux reference is then selected as 
\begin{align}
	\abspsisref = \min(\abspsis^\mtpa, \psismax)
\end{align}
Here, $\psismax$ is the voltage-limited maximum flux magnitude, 
\begin{equation} \label{eq:psismax}
    \psismax = \frac{\ku \udc}{\sqrt{3} |\omegam|}
\end{equation}
where $\udc$ is the DC-bus voltage and $\ku$ is the DC-bus utilization factor. Thus, field weakening is activated when the voltage-limited maximum flux becomes smaller than the MTPA flux. The measured speed can be replaced with the speed estimate in sensorless control or with the (ramped) speed reference in V/Hz control. 

For the selected flux reference, the MTPV and current-limit lookup tables provide the corresponding torque limits,
\begin{align}
	\TM^\mtpv &= \Gmtpv(\abspsisref) \\
	\TM^\clim &= \Glim(\abspsisref) 
\end{align}
The torque reference is limited using these torque limits as
\begin{align} \label{eq:TMreflim}
	\TMreflim = \min\left( |\TMref|, \kmtpv\TM^\mtpv, \TM^\clim \right) \SIGN(\TMref)
\end{align}
where $\kmtpv$ is the MTPV margin. Fig.~\ref{fig:mtpa_mtpv_for_ctrl} shows the precomputed MTPV and current-limit tables for the example machine.

For flux-vector control and V/Hz control, the reference generator outputs the flux reference $\abspsisref$ and the limited torque reference $\TMreflim$. For current-vector control, these two scalar references are converted into a current-reference vector using
\begin{equation}
	\isref = \Gcur(\abspsisref, \TMreflim)
\end{equation}
Depending on the implementation, $\Gcur$ requires one or two 2D lookup tables~\cite{Awa2018}.

The lookup tables are typically generated for positive torque. For negative torque operation, the signs of the limited torque reference and the q-axis current reference are reversed, as shown in Fig.~\ref{fig:ref_gen}.

\begin{table}[t] \centering
\caption{Data of the Four-Pole PM Synchronous Reluctance Machine} \label{tab:machine}
\begin{tabular}{lcc}
\toprule
\textit{Rated values} & \\
\quad Voltage (line-to-neutral, peak value) & $\sqrt{2/3}\cdot 460$ V & 1.00 p.u. \\
\quad Current (peak value) & $\sqrt{2} \cdot 8.8$ A & 1.00 p.u. \\
\quad Frequency & 60 Hz & 1.00 p.u. \\
\quad Speed & 1\,800 r/min & 1.00 p.u. \\
\quad Power & 5.6 kW & 0.80 p.u. \\
\quad Torque & 29.7 Nm & 0.80 p.u. \\
\midrule
\textit{Parameters} & \\
\quad Stator resistance $\Rs$ & 0.63 $\Omega$ & 0.02 p.u. \\
\quad Magnetic model & Fig.~\ref{fig:sat1} & \\ 
\bottomrule
\end{tabular}
\end{table}

\subsection{Proposed Online Method}
As shown in Fig.~\ref{fig:ref_gen}, the proposed online reference-generation method has the same basic signal structure as the conventional lookup-table-based method, but the static lookup tables are replaced by dynamic tracking subsystems. The subsystems $\Gmtpa$, $\Gmtpv$, and $\Glim$ track the MTPA condition, the MTPV limit, and the current limit, respectively. For current-vector control, the additional subsystem $\Gcur$ generates the current reference.

Compared to~\cite{Har2025}, the tracking laws are redesigned so that the internal tracking errors follow exact, decoupled first-order dynamics. This makes the tracking bandwidths easy to tune. The derivations are given in the Appendix. All tracking laws are evaluated at the current state of the corresponding subsystem. For example, the torque of the MTPA subsystem is evaluated as $\TM^\mtpa = \TM(\is^\mtpa)$ using \eqref{eq:tauM}. The tracking laws use only the forward flux map $\psis(\is)$. Thus, the inverse current map $\is(\psis)$ is not needed.

To simplify the notation in the following tracking laws, $\TMref$ denotes the nonnegative torque input applied to the corresponding tracking subsystem. In the complete reference-generation architecture, the MTPA subsystem uses $|\TMreflim|$ as this input. The sign of the torque is handled separately, as shown in Fig.~\ref{fig:ref_gen}.

\subsubsection{MTPA}
The MTPA current vector $\is^\mtpa$ is defined as the stator current that produces the torque reference $\TMref$ with minimum current magnitude. The dynamic subsystem $\Gmtpa$ receives the torque reference as its input. The tracking is driven by two errors: the torque error $\TMref - \TM^\mtpa$ and the MTPA condition error $\cgamma^\mtpa$. The tracking law is 
\begin{align} \label{eq:mtpa_law}
    \frac{\D \is^\mtpa}{\D t} = \alphamtpa \frac{ (\TMref - \TM^\mtpa)\J\phia^\mtpa + \cgamma^\mtpa\J\psia^\mtpa}{{(\psia^\mtpa)}^\T \J\phia^\mtpa}
\end{align}
where $\alphamtpa$ is the tracking bandwidth. The state $\is^\mtpa$ is mapped through the forward flux map to obtain the MTPA flux magnitude $\abspsis^\mtpa = \|\psis(\is^\mtpa)\|$. This signal is the primary output of $\Gmtpa$. If necessary, other optimal quantities can be readily extracted, as they depend statically on the state $\is^\mtpa$.

\subsubsection{MTPV}
The dynamic subsystem $\Gmtpv$ receives the flux reference $\abspsisref$ as its input. The tracking of the MTPV current state $\is^\mtpv$ is driven by the flux-magnitude error $\abspsisref - \abspsis^\mtpv$ and the MTPV condition error $\cdelta^\mtpv$. The tracking law is
\begin{align} \label{eq:mtpv_law}
    \frac{\D \is^\mtpv}{\D t} = \alphamtpv \frac{ (\abspsisref - \abspsis^\mtpv)\J\phie^\mtpv + \cdelta^\mtpv \J \lvect^\mtpv }{{(\lvect^\mtpv)}^\T\J\phie^\mtpv}
\end{align} 
where $\alphamtpv$ is the tracking bandwidth. Although the MTPV condition is naturally expressed in flux coordinates, the current is used as the state variable here. This allows all required quantities to be evaluated using the forward flux map and avoids the inverse current map. The subsystem output is the MTPV torque limit $\TM^\mtpv = \TM(\is^\mtpv)$.

\begin{table}[t]
\centering
\caption{Control System Parameters} \label{tab:ctrl}
\begin{tabular}{lc}
\toprule
\textit{Flux-vector control} & \\
\quad Flux-control bandwidth $\alpha_\psiup$ & $2\piup \cdot 100$ rad/s \\
\quad Torque-control bandwidth $\alpha_\tauup$ & $2\piup \cdot 100$ rad/s \\
\quad Speed-estimation bandwidth $\alpha_\mathrm{o}$ & $2\piup \cdot 8$ rad/s \\
\quad Voltage utilization factor $\ku$ & 0.85 \\
\quad MTPV margin $\kmtpv$ & 0.70 \\
\quad Maximum current $\ismax$ & 2 p.u. \\
\midrule
\textit{Speed control} & \\
\quad Speed-control bandwidth $\alpha_\mathrm{s}$ & $2\piup \cdot 4$ rad/s \\
\quad Total inertia $J$ & 0.05 kgm$^2$ \\
\bottomrule
\end{tabular}
\end{table}

\subsubsection{Current Limit}
The subsystem $\Glim$ also receives the flux reference $\abspsisref$ as its input. Its state $\is^\clim$ is constrained to the current-limit contour. The tracking is driven by the flux-magnitude error $\abspsisref - \abspsis^\clim$, i.e.,
\begin{align} \label{eq:clim_law}
    \frac{\D \is^\clim}{\D t} = \alphalim \frac{(\abspsisref - \abspsis^\clim) \J\is^\clim}{{(\lvect^\clim)}^\T \J \is^\clim}
\end{align}
where $\alphalim$ is the tracking bandwidth. The state is then used to calculate the current-limit torque $\TM^\clim = \TM(\is^\clim)$. As shown in the Appendix, this formulation keeps the subsystem state on the prescribed current-limit contour.  

\subsubsection{Current Reference}
The subsystem $\Gcur$ is needed only when the reference generator must provide current references, as in current-vector control. It receives the torque reference $\TMref$ and the flux reference $\abspsisref$ as inputs. The tracking of the current state $\is^\cc$ is driven by the torque error $\TMref - \TM^\cc$ and the flux-magnitude error $\abspsisref - \abspsis^\cc$. The tracking law is
\begin{align} \label{eq:cur_law}
    \frac{\D \is^\cc}{\D t} = \alphacc \frac{(\TMref - \TM^\cc)\J\lvect^\cc - (\abspsisref - \abspsis^\cc)\J\psia^\cc }{(\psia^\cc)^\T \J \lvect^\cc}
\end{align}
where $\alphacc$ is the tracking bandwidth. The subsystem output is the current reference supplied to the current controller, i.e., $\isref = \is^\cc$. For negative torque, the same sign convention as in Fig.~\ref{fig:ref_gen} is used: the sign of the q-axis current reference is reversed.

The subsystem $\Gcur$ can also be combined with static 1D lookup tables for the MTPA, MTPV, and current limits. In this hybrid implementation, $\Gcur$ replaces the conventional 2D current-reference lookup tables, so only 1D lookup tables are needed.

\begin{figure*}[t] 
    \centering
    \subfigure[]{\includegraphics[width=0.325\textwidth]{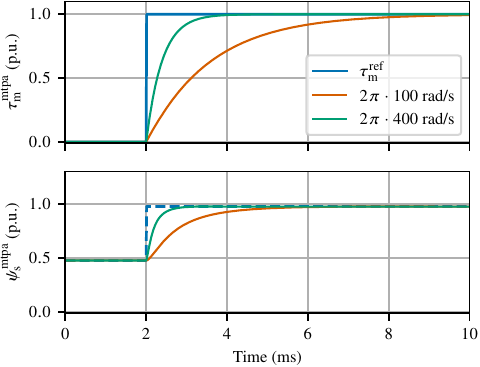}} \hfill
    \subfigure[]{\includegraphics[width=0.325\textwidth]{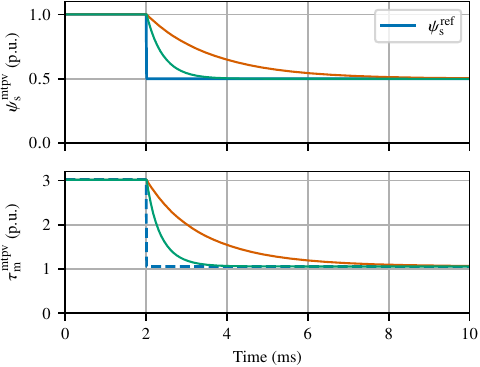}} \hfill
    \subfigure[]{\includegraphics[width=0.325\textwidth]{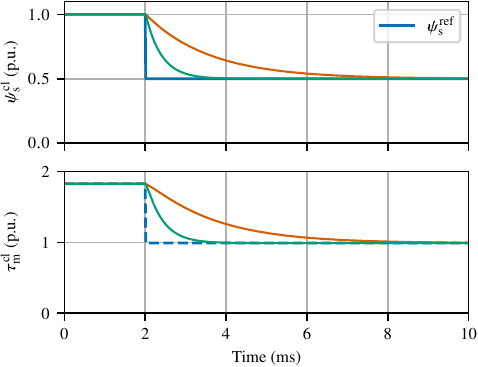}} 
    \caption{Simulated step responses for the proposed method: (a) MTPA; (b) MTPV; (c) current limit. The tracking bandwidths are $2\piup \cdot$ 100 rad/s (orange) and $2\piup \cdot 400$ rad/s (green). The dashed lines represent the conventional lookup-table method.} \label{fig:sim_tau_flux}
\end{figure*}

\subsection{Implementation Aspects} \label{sec:implementation}
Although the tracking laws are formulated in continuous time, they can be implemented digitally using forward Euler discretization. At each sampling instant, the right-hand side of each tracking law is evaluated at the current subsystem state, and the state is advanced by one sampling step.

The transient response of the generated references is set by the bandwidths $\alphamtpa$, $\alphamtpv$, $\alphalim$, and $\alphacc$. In practice, a common bandwidth can be used for all subsystems. This bandwidth should be selected well below the sampling frequency to maintain a numerical margin. 

To reduce the computational load, the highest-order gradient terms in the MTPA and MTPV gradients can be omitted. As shown in the Appendix, the resulting error dynamics remain lower triangular. Thus, the primary tracking errors retain their first-order dynamics, and the steady-state optimality conditions are preserved. However, this approximation introduces some cross-coupling during transients.

The dynamic states are initialized to avoid startup transients. The reference generator can be started from the zero-torque operating point, i.e., with $\TMref = 0$. The initial flux reference is then the zero-torque MTPA flux, which for a PM machine corresponds to the no-load flux linkage. The initial subsystem states are set consistently with this operating point. Once the tracking laws are enabled, the states evolve toward the active references.

\begin{figure}[t]
        \centering
        \includegraphics[width=0.325\textwidth]{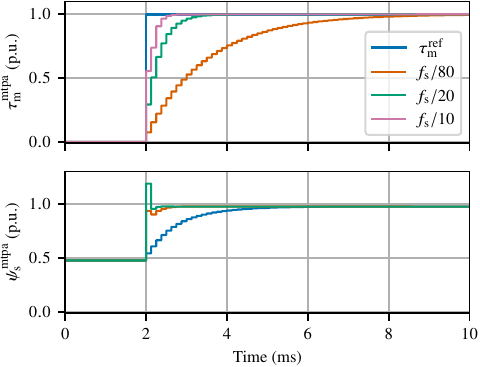}
        \caption{Simulated discrete-time MTPA step response for the proposed method. The tracking bandwidths are $f_\mathrm{s}/80$, $f_\mathrm{s}/20$, and $f_\mathrm{s}/10$.} \label{fig:mtpa_step_discrete}
\end{figure}

\section{Results} \label{sec:results}
The 5.6-kW four-pole PM synchronous reluctance machine (ABB Baldor ECS101M0H7EF4) is used to validate the proposed reference-generation method. The rated values and parameters are given in Table~\ref{tab:machine}. The flux maps were measured using a constant-speed test~\cite{Arm2013} and fitted to an analytical saturation model~\cite{Lel2024}, as shown in Fig.~\ref{fig:sat1}. 

In the simulations, the reference-generation algorithm is evaluated independently of the control system and machine model. This allows the intrinsic tracking behavior of the algorithm to be studied without the influence of the real-time control system or hardware. By contrast, the experiments evaluate the algorithm as part of the complete real-time closed-loop drive system.

\subsection{Simulation Results} \label{sec:sim}
To validate the analytically derived tracking dynamics and to demonstrate that the transient response can be tuned directly, the system responses to step changes in the torque and flux references are first simulated in continuous time. In Fig.~\ref{fig:sim_tau_flux}(a), the torque reference $\TMreflim$ is stepped from 0 to 1 p.u. at $t =$ 2 ms to demonstrate the MTPA tracking response governed by~\eqref{eq:mtpa_law}. In Figs.~\ref{fig:sim_tau_flux}(b) and (c), the flux reference $\abspsisref$ is stepped from 1 p.u. to 0.5 p.u. to evaluate the current-limit and MTPV-limit tracking responses, determined by~\eqref{eq:clim_law} and~\eqref{eq:mtpv_law}, respectively. Both scenarios are simulated using two different tracking bandwidths ($\alpha = 2\piup \cdot 80$~rad/s and $\alpha = 2\piup \cdot 400$~rad/s, where $\alpha = \alphamtpv = \alphalim = \alphacc$) and compared with the conventional lookup-table method. 

The forward Euler discretization is simulated to assess the discrete-time stability of the algorithm. Fig.~\ref{fig:mtpa_step_discrete} shows the MTPA tracking response when the bandwidth is varied from $f_\mathrm{s}/80$ to $f_\mathrm{s}/10$, where $f_\mathrm{s}$ is the sampling frequency. The method remains numerically stable until the bandwidth approaches $f_\mathrm{s}/5$.

\begin{figure}[t]
        \centering
        \includegraphics[width=0.36\textwidth]{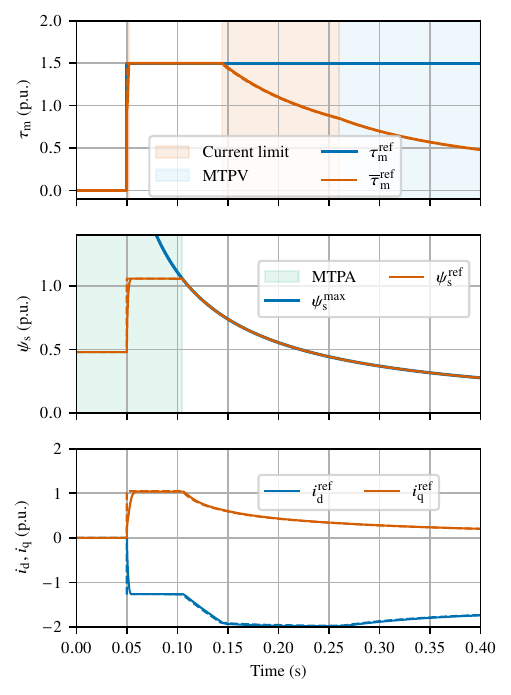}
        \caption{Simulation results for the proposed method during a full-range speed ramp. The torque reference is stepped from 0 to 1.5~p.u. at $t = 0.05$~s. The algorithm successfully generates continuous current references while strictly respecting the MTPA trajectory, the current limit, field weakening, and the MTPV limit. The dashed lines represent the conventional lookup-table method.} \label{fig:sim_full_dynamics}
\end{figure}

To verify operation across all regions, a full-range simulation is carried out using a uniform tracking bandwidth of $2\piup \cdot 100$~rad/s and a constant DC-link voltage $\udc =$ 540~V. In Fig.~\ref{fig:sim_full_dynamics}, the torque reference is stepped from 0 to 1.5 p.u., while the rotor speed $\omegam$ is ramped from 0 to 3 p.u. Due to the finite DC-link voltage, the maximum allowable stator flux magnitude \eqref{eq:psismax} decreases as the speed increases. When this limit falls below the MTPA flux $\abspsis^\mtpa$, the active flux reference $\abspsisref$ is reduced, initiating field weakening. During this wide-speed-range transition, the current references are continuously generated, governed by the current-tracking dynamics \eqref{eq:cur_law}. As mentioned earlier, to isolate the intrinsic tracking behavior of the reference generator, these simulations are conducted independently of the closed-loop control system.

The results demonstrate that the proposed method successfully tracks the MTPA locus, the current limit, the field-weakening region, and the MTPV limit. While the conventional lookup-table method provides reference signals instantaneously, the proposed approach tracks the same optimal loci with transients determined by the selected bandwidths. In all simulated scenarios, the steady-state references generated by the proposed method exactly match those obtained with the conventional lookup-table approach.

\subsection{Experimental Results} \label{sec:exp}
Experimental tests were conducted to evaluate the proposed method using the 5.6-kW PM synchronous reluctance machine drive. The control system employs the flux-vector control scheme from \cite{Tii2024} together with the algebraic saturation model \cite{Lel2024}, fitted to the measured data shown in Fig.~\ref{fig:sat1}. The same saturation model is used consistently throughout the control system. The parameters of the control system are given in Table~\ref{tab:ctrl}.

\begin{figure}[t] 
    \centering
        \centering
        \subfigure[]{\includegraphics[width=0.41\textwidth]{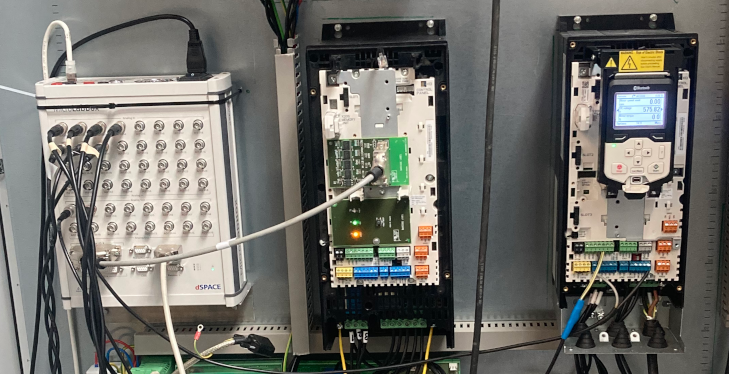}} 
        \subfigure[]{\includegraphics[width=0.41\textwidth]{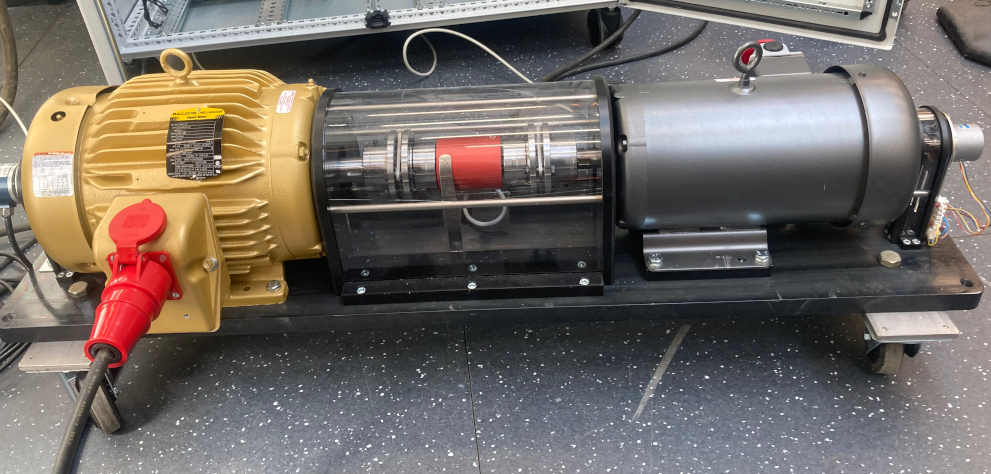}} 
        \caption{Experimental setup: (a) dSPACE MicroLabBox (left), inverter for the test machine (center), and inverter for the load machine (right); (b) load machine (left) and test machine (right).} \label{fig:lab_setup}
\end{figure}

Fig.~\ref{fig:lab_setup} shows the laboratory setup. The test machine is supplied by a two-level voltage source inverter, while the control algorithm is executed on a dSPACE MicroLabBox platform. The switching signals generated by the dSPACE system are routed directly to the inverter. A switching frequency of 8~kHz, combined with a double-sampling technique, is used. As a mechanical load, a 5.6-kW induction machine is operated in torque-control mode via a dedicated inverter. Rotor speed is measured using a resolver, and stator currents are measured using Hall-effect sensors.

\begin{figure*}[t] 
    \centering
    \subfigure[]{\includegraphics[width=0.37\textwidth, trim=0cm .4cm 0cm .3cm]{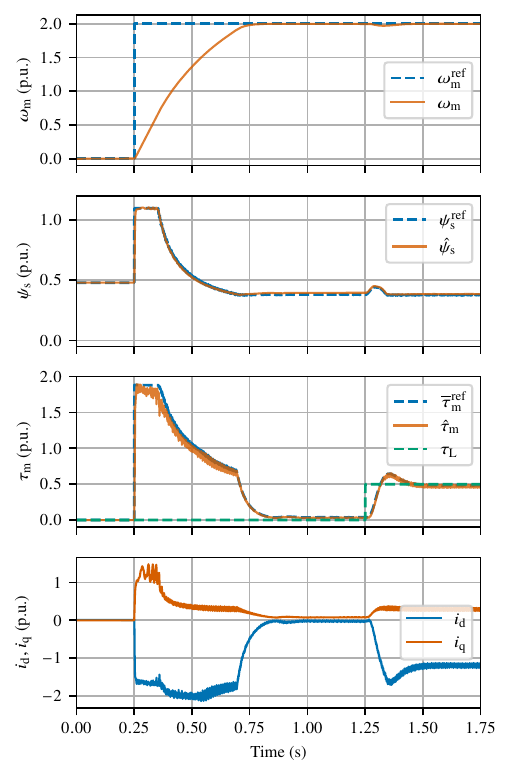} \label{fig:meas_lookup}}
    \qquad\qquad\quad
    \subfigure[]{\includegraphics[width=0.37\textwidth, trim=0cm .4cm 0cm .3cm]{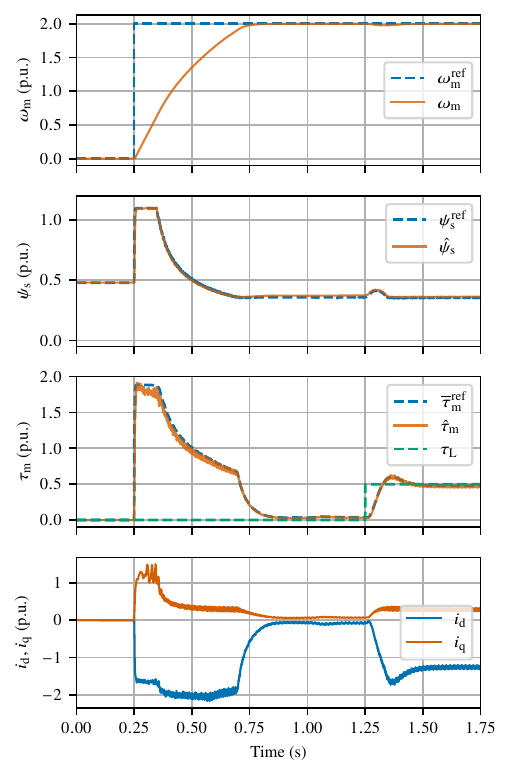} \label{fig:meas_proposed}}
    \caption{Experimental results for the speed-reference step to twice the nominal speed and the subsequent load-torque step: (a) conventional lookup-table method; (b) proposed method.} \label{fig:experimental_results}
\end{figure*}

Figs.~\ref{fig:meas_lookup} and~\ref{fig:meas_proposed} show the experimental results for the conventional lookup-table method and the proposed method, respectively. The test sequence consists of a speed-reference step from 0 to 2~p.u. (3\,600~r/min) at $t = 0.25$~s, followed by a load-torque step at $t = 1.25$~s. During this sequence, the drive traverses all operating regions: the MTPA locus, the current limit, the field-weakening region, and the MTPV limit. Due to the defined margins and the relatively low DC-link voltage, the system naturally enters the MTPV region.

The results demonstrate that the proposed method successfully handles all operating regions, achieving steady-state and dynamic performance on par with the conventional lookup-table method. Furthermore, the experiments confirm that despite replacing static tables with continuous nonlinear tracking laws, the dynamic reference generator executes reliably in real time on the dSPACE hardware, alongside the primary control loops operating at a 16-kHz sampling rate.

Because the reference generator operates in a feedforward manner, its transient dynamics are determined directly by its bandwidth parameters. It remains structurally separate from the feedback control system and avoids creating an additional feedback path. This separation simplifies tuning and preserves the intended behavior of the main feedback loops.

\section{Conclusions} \label{sec:conclusion}
This paper presented an online feedforward reference-generation method for saturable synchronous machines. The method dynamically generates optimal references without precomputed lookup tables and tracks the MTPA trajectory while respecting the current limit, the field-weakening operation, and the MTPV limit over the full speed range. A key contribution is the formulation of tracking laws with exact first-order error dynamics, which simplifies tuning of the method. Moreover, the method uses only the forward flux map and does not require an inverse current map. 

Because the reference generator operates in a feedforward manner, its dynamics remain structurally separate from the inner feedback control loops. Furthermore, the reference signals are free from measurement noise. The method was validated through simulations and laboratory experiments on a 5.6-kW PM synchronous reluctance machine drive. The results show that the proposed method achieves steady-state and transient performance comparable to conventional lookup-table-based methods while avoiding precomputed reference tables.

\appendix[Derivations of Tracking Error Dynamics]
Throughout this appendix, the error dynamics of the proposed tracking subsystems are derived using a unified mathematical structure. By applying the respective tracking laws and the chain rule, $\D y/\D t = (\partial y/\partial \is)^\T (\D \is/\D t)$, the resulting error dynamics rely on the skew-symmetric properties of the matrix $\J$. In particular, for any vectors $\mathbf{v}$ and $\mathbf{w}$, the identity $\mathbf{v}^\T\J\mathbf{v} = 0$ eliminates the cross terms, yielding exactly decoupled first-order dynamics, while $\mathbf{v}^\T\J\mathbf{w} = -\mathbf{w}^\T\J\mathbf{v}$ is used to cancel the denominators.

\subsection{MTPA}
Applying the MTPA tracking law \eqref{eq:mtpa_law} and the above skew-symmetric identities, the tracking dynamics for the MTPA torque and condition function become
\begin{align} \nonumber
    \frac{\D \TM^\mtpa}{\D t} \!&=\! \frac{\alphamtpa {(\psia^\mtpa)}^\T}{{(\psia^\mtpa)}^\T\J\phia^\mtpa} \! \left[ (\TMref \!-\! \TM^\mtpa)\J\phia^\mtpa + \cgamma^\mtpa\J\psia^\mtpa \right] \\ 
    \!&=\! \alphamtpa (\TMref - \TM^\mtpa) \\ \nonumber
    \frac{\D \cgamma^\mtpa}{\D t} \!&=\! \frac{\alphamtpa {(\phia^\mtpa)}^\T}{{(\psia^\mtpa)}^\T\J\phia^\mtpa} \! \left[ (\TMref \!-\! \TM^\mtpa)\J\phia^\mtpa + \cgamma^\mtpa\J\psia^\mtpa \right] \\ 
    \!&=\! -\alphamtpa \cgamma^\mtpa
\end{align}

\subsection{MTPV}
Similarly, evaluating the MTPV tracking law \eqref{eq:mtpv_law} yields the decoupled dynamics for the MTPV flux and condition function,
\begin{align} \nonumber
    \frac{\D \abspsis^\mtpv}{\D t} \!&=\! \frac{\alphamtpv {(\lvect^\mtpv)}^\T}{{(\lvect^\mtpv)}^\T\J\phie^\mtpv} \! \left[ (\abspsisref \!-\! \abspsis^\mtpv)\J\phie^\mtpv + \cdelta^\mtpv\J\lvect^\mtpv \right] \\ 
    \!&=\! \alphamtpv (\abspsisref - \abspsis^\mtpv) \\ \nonumber
    \frac{\D \cdelta^\mtpv}{\D t} \!&=\! \frac{\alphamtpv {(\phie^\mtpv)}^\T}{{(\lvect^\mtpv)}^{\T}\J\phie^{\mtpv}} \! \left[ (\abspsisref \!-\! \abspsis^\mtpv)\J\phie^\mtpv + \cdelta^\mtpv\J\lvect^\mtpv \right] \\ 
    \!&=\! -\alphamtpv \cdelta^\mtpv
\end{align}

\subsection{Current Limit}
Applying the current limit tracking law \eqref{eq:clim_law}, the flux linkage magnitude tracks its reference with exact first-order dynamics:
\begin{align} \nonumber
    \frac{\D \abspsis^\clim}{\D t} &= \frac{\alphalim {(\lvect^{\clim})}^{\T}}{{(\lvect^{\clim})}^{\T}\J\is^\clim} (\abspsisref - \abspsis^{\clim}) \J\is^\clim \\ 
    &= \alphalim (\abspsisref - \abspsis^\clim)
\end{align}
Furthermore, evaluating the time derivative of the squared current magnitude shows that the tracking state remains on the maximum-current-limit contour:
\begin{align}
    \frac{1}{2} \frac{\D \|\is^\clim\|^2}{\D t} &= \frac{\alphalim (\abspsisref - \abspsis^\clim)}{{(\lvect^\clim)}^\T\J\is^\clim} {(\is^\clim)}^\T \J\is^\clim = 0
\end{align}

\subsection{Current Reference}
Under the current-reference tracking law \eqref{eq:cur_law}, the torque and flux magnitude tracking dynamics are similarly decoupled:
\begin{align} \nonumber
    \frac{\D \TM^\cc}{\D t} &= \frac{\alphacc {(\psia^\cc)}^\T}{{(\psia^\cc)}^\T \J \lvect^\cc} \left[ (\TMref - \TM^\cc) \J \lvect^\cc - (\abspsisref - \abspsis^\cc) \J \psia^\cc \right] \\ 
    &= \alphacc (\TMref - \TM^\cc) \\ \nonumber
    \frac{\D \abspsis^\cc}{\D t} &= - \alphacc (\abspsisref - \abspsis^\cc) \frac{{(\lvect^\cc)}^\T \J \psia^\cc}{{(\psia^\cc)}^\T \J \lvect^\cc} \\ 
    &= \alphacc (\abspsisref - \abspsis^\cc)
\end{align}

\subsection{Truncated Gradient Terms}
Let $\tildephia$ denote the approximate gradient in which the computationally intensive partial-derivative terms are omitted. Executing the MTPA tracking law \eqref{eq:mtpa_law} with $\tildephia$ yields the modified error dynamics
\begin{align}
    \frac{\D}{\D t} \begin{bmatrix} e_\tauup \\ \cgamma \end{bmatrix} &= \begin{bmatrix} -\alphamtpa & 0 \\[6pt] \alphamtpa \frac{{(\phia^\mtpa - \tildephia)}^\T \J \tildephia}{\psia^\T \J \tildephia} & -\alphamtpa \frac{\psia^\T \J \phia}{\psia^\T \J \tildephia} \end{bmatrix} \begin{bmatrix} e_\tauup \\ \cgamma \end{bmatrix}
\end{align}
where $e_\tauup = \TMref - \TM^\mtpa$. The MTPV implementation follows the same logic. Because the identities $\psia^\T\J\psia = 0$ and ${(\lvect^\mtpv)}^\T\J\lvect^\mtpv = 0$ still hold, the resulting system matrices remain lower-triangular. Thus, the main tracking error $e_\tauup$ follows its first-order dynamics, and the steady-state optimal point is unchanged, while the condition-error dynamics may be driven by one-way coupling from the main tracking error.

\IEEEtriggeratref{4}

\begin{IEEEbiography}[{\includegraphics[width=1in,height=1.25in,clip,keepaspectratio]{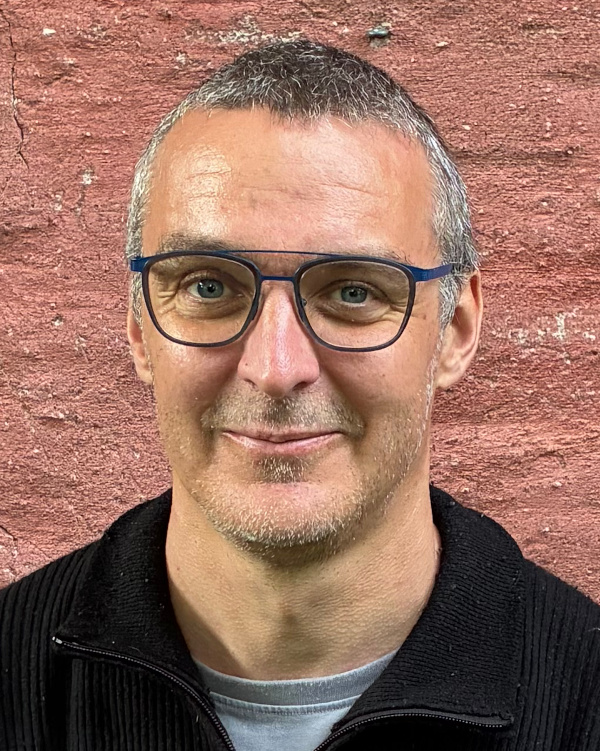}}]
{Mikko Sarén} received the B.Sc.(Tech.) and M.Sc.(Tech.) degrees in electrical engineering in 2023 and 2025, respectively, from Aalto University, Espoo, Finland, where he is currently working toward the D.Sc.(Tech.) degree in electrical engineering.

His research interests include the control of electric motor drives.
\end{IEEEbiography}

\begin{IEEEbiography}[{\includegraphics[width=1in,height=1.25in,clip,keepaspectratio]{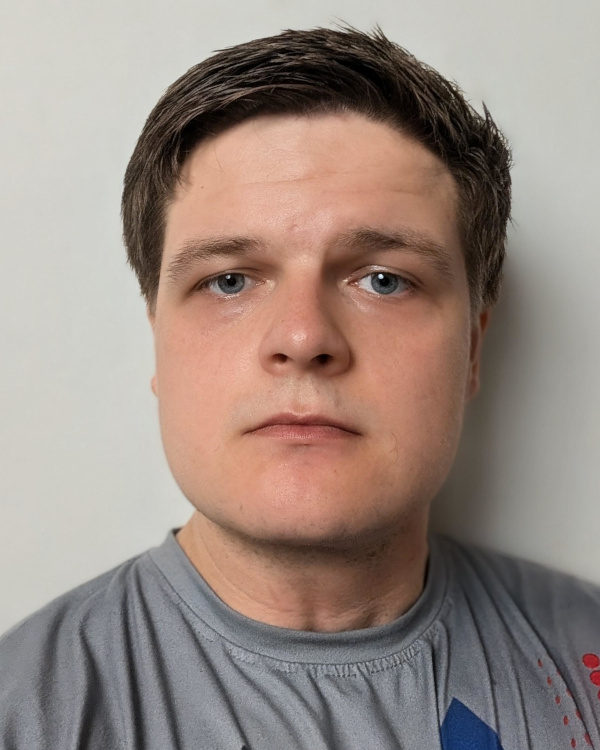}}]
{Hannu Hartikainen} received the B.Sc. degree in electrical engineering from the Metropolia University of Applied Sciences, Helsinki, Finland, in 2020, and the M.Sc.(Tech.) degree in electrical engineering in 2022 from Aalto University, Espoo, Finland, where he is currently working toward the D.Sc.(Tech.) degree in electrical engineering.

His research interests include the control of electric machines with an LC filter.
\end{IEEEbiography}

\begin{IEEEbiography}[{\includegraphics[width=1in,height=1.25in,clip,keepaspectratio]{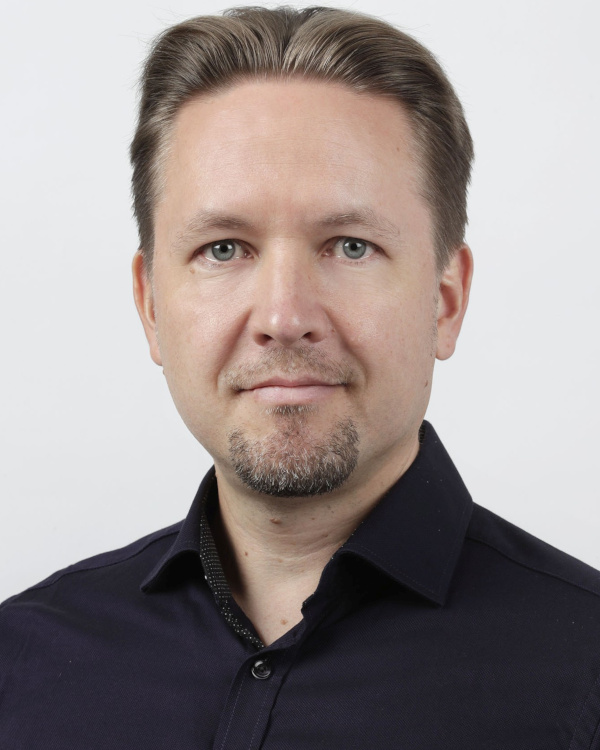}}]
{Antti Piippo} received the M.Sc.(Eng.) and D.Sc.(Tech.) degrees in electrical engineering from the Helsinki University of Technology, Espoo, Finland, in 2003 and 2008, respectively.
 
He is currently an R\&D Executive Engineer with ABB Oy, Drives, Helsinki, Finland. His main research interests include the control of electric drives. 
\end{IEEEbiography}

\begin{IEEEbiography}[{\includegraphics[width=1in,height=1.25in,clip,keepaspectratio]{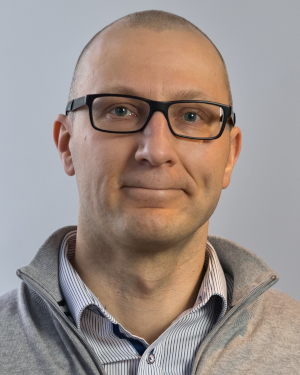}}]
{Marko Hinkkanen} (M'06--SM'13--F'23) received the M.Sc.(Eng.) and D.Sc.(Tech.) degrees in electrical engineering from the Helsinki University of Technology, Espoo, Finland, in 2000 and 2004, respectively.

He is currently a Full Professor with the School of Electrical Engineering, Aalto University, Espoo, Finland. His research interests include control systems, physics-informed machine learning, electric machine drives, and power converters.

Dr. Hinkkanen was the recipient of several paper awards, including the 2016 International Conference on Electrical Machines (ICEM) Brian J. Chalmers Best Paper Award, and the 2016 and 2018 IEEE Industry Applications Society Industrial Drives Committee Best Paper Awards. He was the corecipient of the 2020 SEMIKRON Innovation Award. He was the General Cochair of the 2018 IEEE 9th International Symposium on Sensorless Control for Electrical Drives (SLED). He is an Associate Editor of \textsc{IEEE Transactions on Power Electronics}.
\end{IEEEbiography}

\end{document}

%% file: vardefs.tex
\DeclareMathOperator{\SIGN}{sign}

\newcommand{\D}{\mathrm{d}}
\newcommand{\T}{\top}               
\newcommand{\J}{\mathbf{J}}

\newcommand{\udc}{u_\mathrm{dc}}
\newcommand{\psisd}{\psi_\mathrm{d}}
\newcommand{\psisq}{\psi_\mathrm{q}}
\newcommand{\abspsis}{\psi_\mathrm{s}}
\newcommand{\isd}{i_\mathrm{d}}
\newcommand{\isq}{i_\mathrm{q}}

\newcommand{\Rs}{R_\mathrm{s}}
\newcommand{\us}{\mathbf{u}_\mathrm{s}}
\newcommand{\is}{\mathbf{i}_\mathrm{s}}
\newcommand{\psis}{\bm{\psiup}_\mathrm{s}}
\newcommand{\omegam}{\omega_\mathrm{m}}
\newcommand{\thetam}{\vartheta_\mathrm{m}}

\newcommand{\Gs}{\boldsymbol{\Gamma}_\mathrm{\!s}}
\newcommand{\Ls}{\mathbf{L}_\mathrm{s}}
\newcommand{\Ld}{L_\mathrm{d}}
\newcommand{\Lq}{L_\mathrm{q}}
\newcommand{\abspsif}{\psi_\mathrm{f}}

\newcommand{\TM}{\tau_\mathrm{m}}
\newcommand{\ia}{\mathbf{i}_\mathrm{a}}
\newcommand{\psia}{\bm{\psiup}_\mathrm{a}}

\newcommand{\phia}{\bm{\phiup}_\gammaup}
\newcommand{\tildephia}{\tilde{\bm{\phiup}}_\gammaup}
\newcommand{\ja}{\mathbf{i}_\mathrm{e}}
\newcommand{\lvect}{\bm{\ell}}
\newcommand{\phie}{\bm{\phiup}_\mathrm{e}}
\newcommand{\cdelta}{e_\mathrm{ia}}
\newcommand{\cgamma}{e_{\psiup\mathrm{a}}}

\newcommand{\abspsisref}{\abspsis^\mathrm{ref}}
\newcommand{\TMref}{\TM^\mathrm{ref}}
\newcommand{\TMreflim}{\overline{\tau}_\mathrm{m}^\mathrm{ref}}
\newcommand{\isref}{\is^\mathrm{ref}}

\newcommand{\mtpa}{\mathrm{mtpa}}
\newcommand{\mtpv}{\mathrm{mtpv}}
\newcommand{\clim}{\mathrm{cl}}
\newcommand{\cc}{\mathrm{cur}}

\newcommand{\Gmtpa}{\Sigma_\mtpa}
\newcommand{\Gmtpv}{\Sigma_\mtpv}
\newcommand{\Glim}{\Sigma_\clim}
\newcommand{\Gcur}{\Sigma_\cc}

\newcommand{\alphamtpa}{\alpha_\mtpa}
\newcommand{\alphamtpv}{\alpha_\mtpv}
\newcommand{\alphalim}{\alpha_\clim}
\newcommand{\alphacc}{\alpha_\cc}

\newcommand{\kmtpv}{k_\mtpv}
\newcommand{\ku}{k_\mathrm{u}}

\newcommand{\psismax}{\abspsis^\mathrm{max}}

\newcommand{\ismax}{i_\mathrm{s}^\mathrm{max}}